# Results from a Prototype Combination TPC Cherenkov Detector with GEM Readout

B. Azmoun, K. Dehmelt, T. K. Hemmick, R. Majka, H. N. Nguyen, M. Phipps, M. L. Purschke, N. Ram, W. Roh, D. Shangase, N. Smirnov, C. Woody, A. Zhang

*Abstract*—A combination Time Projection Chamber-Cherenkov prototype detector has been developed as part of the Detector R&D Program for a future Electron Ion Collider. The prototype was tested at the Fermilab test beam facility to provide a proof of principle to demonstrate that the detector is able to measure particle tracks and provide particle identification information within a common detector volume. The TPC portion consists of a $10 \times 10 \times 10 \text{cm}^3$ field cage, which delivers charge from tracks to a $10 \times 10 \text{cm}^2$ quadruple GEM readout. Tracks are reconstructed using charge and timing information from clusters collected on an array of $2 \times 10 \text{mm}^2$ zigzag pads. The Cherenkov portion consists of a $10 \times 10 \text{cm}^2$ readout plane segmented into 3x3 square pads, also coupled to a quadruple GEM. As tracks pass though the drift volume of the TPC, the generated Cherenkov light is able to escape through sparsely arranged wires making up one side of the field cage, facing the CsI photocathode of the Cherenkov detector. The Cherenkov detector is thus operated in a windowless, proximity focused configuration for high efficiency. Pure $CF_4$ is used as the working gas for both detector components, mainly due to its transparency into the deep UV, as well as its high $N_0$. Results from the beam test, as well as results on its particle id capabilities will be discussed.

## I. INTRODUCTION

The ability to track charged particles and provide particle identification (pID) in the same detector offers considerable advantages in terms of efficiency, reducing material and multiple scattering, utilization of space inside a spectrometer, and minimizing cost. A multipurpose detector has been studied that combines the tracking features of a Time Projection Chamber (TPC) with additional pID from a threshold Cherenkov detector. This work is part of a Detector R&D Program for a future Electron Ion Collider that is being planned to be built at either Brookhaven National Lab (eRHIC) or Thomas Jefferson National Lab (MEIC) [1, 2]. An EIC would collide beams of electrons with protons and heavy ions at high energies in order to study nucleon structure and QCD over a broad range of x and $Q^2$. A large multipurpose spectrometer would be used to measure deep inelastic electron scattering over a wide range of rapidity and solid angle. The tracking system for the central detector could consist of a TPC and a precision vertex detector. The combined TPC-Cherenkov detector described here could be used to provide both tracking and pID information for measuring the scattered electron and separating it from hadrons produced in the central region.

We have constructed a GEM-based prototype TPC-Cherenkov detector (TPCC) that combines the tracking features of a Time Projection Chamber (TPC) and a threshold Cherenkov detector for pID in a common detector volume. This prototype was tested at the Fermilab Test Beam Facility (FTBF) to provide a proof of principle for the viability of this detector concept. In this paper we report on both the tacking performance and on the pID efficiency for this prototype.

## II. EXPERIMENTAL SETUP

A model of the TPCC prototype is shown in Fig. 1, with the particle beam entering the detector volume from the left side. In order to achieve efficient pID in a compact design, the Cherenkov yield must be maximized. For this reason, the detector chamber is filled with high purity $CF_4$, which is a radiator capable of very high Cherenkov yields. At the same time, $CF_4$ acts as a suitable operating gas for the GEM's and is a very fast drift gas for the TPC. The primary ionization created by the passage of charged particles through the drift volume is drifted downward by the drift field toward the TPC GEM where the signal is amplified and read out. At the same time, the generated Cherenkov light passes through the transparent side of the field cage closest to the Cherenkov portion of the prototype and impinges a photosensitive GEM detector, which provides a simultaneous electron trigger.

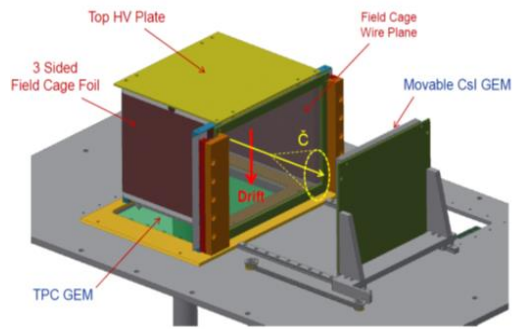

**Fig. 1** Engineering model of the TPCC prototype.

### A. TPC detector

The TPC portion of the prototype consists of a standard quadruple GEM detector coupled to a field cage with a drift volume roughly 10cm x 10cm x 10cm. As shown in Fig. 2, the readout plane is segmented into fifty 10mm long zigzag shaped

Manuscript submitted on December 21, 2018. This work was supported in part by the U.S. Department of Energy under Prime Contract No. DE-SC0012704.

B.Azmoun, M.L.Purschke, C.Woody and A. Zhang are with Brookhaven National Laboratory, Upton, NY.
K.Dehmelt, T.K.Hemmick, H.N.Nguyen, N.Ram, W.Roh, and D.Shangase are with Stony Brook University, Stony Brook, NY.
M.Phipps is with the University of Illinois at Urbana Champaign
R.Majka and N.Smirnov are with Yale University, New Haven, CT.

anodes, arranged in ten pad-rows for a total of 500 pads, which make up an active area of 10cm x 10cm. Neighboring zigzags are interleaved with one another and have a 2mm pitch along the position sensitive coordinate ($X$). As shown in another TPC study [3], the purpose of the zigzag electrode design is to enhance charge sharing along the sensitive coordinate, where good position resolution must be achieved with a limited number of electronics channels [4].

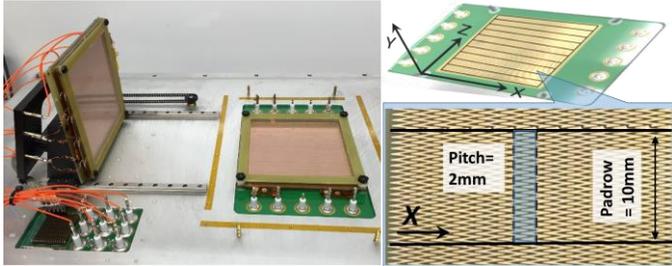

**Fig. 2 Expanded view of the TPCC prototype without the TPC field cage installed. The TPC zigzag readout PCB and the respective coordinate system are also shown.**

The field cage comprises a kapton foil with twenty-five parallel field forming electrode strips on one side, with a pitch of 4mm. On the other side of the foil, similar sized "mirror" strips are staggered by half the pitch, which allows for a finer field gradient for the purpose of improving the field uniformity in the drift volume and to also minimize field punch-through to the exterior. The strips are each 3.7mm wide with a 300μm gap between them to prevent sparking. The side of the field cage facing the Cherenkov detector is effectively made transparent to the generated light within the drift volume by replacing the foil strips with thin, 75μm diameter wires, spaced 1mm apart. Each group of 4 wires receives the same potential as the foil strips such that a fixed potential is maintained at each transverse slice along the drift direction. A picture of the field cage with and without the wire electrodes is shown in Fig. 3.

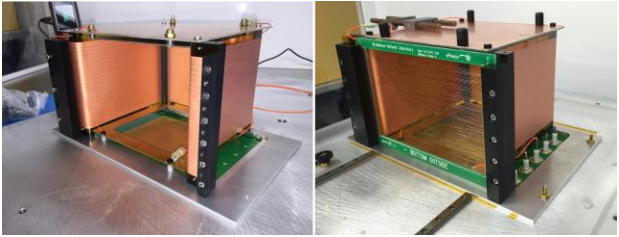

**Fig. 3 The TPC field cage strip foil with and without the plane of wire electrodes.**

At the top of the field cage the field is terminated in a planar electrode or "top plate". At the bottom, a similar "bottom plate" with the center cut out is used to help form the field between the field cage and the GEM stack.

An average electric field of 400V/cm was established in the drift volume by applying a potential of roughly 4kV between the top and bottom plates of the field cage, which are separated by about 10cm. At this field, a voltage may be applied to the top plate to achieve a drift velocity for $CF_4$ of about 7.5μm/ns, which is relatively fast and suitable for the available time window of our front end electronics. A passive voltage divider, shown in Fig. 4 provided a voltage drop of about 160V across each strip electrode of the inner foil of the field cage. Each inner strip is also connected to an outer "mirror" strip behind it such that two staggered strips are at the same potential. The drift gap electrode of the quadruple GEM stack is made of a fine planar mesh located a few mm below the bottom plate. The magnitude of the field between the bottom plate and the mesh was tuned by employing a second power supply to fix the potential of the bottom plate, which is centered on the bottom-most inner strip. Two more power supplies are used to power the GEM stack: one is used to set the potential of the mesh to achieve the desired drift gap field and a second is used to power the GEMs. As seen in the figure, HV is distributed to each GEM by a second voltage divider, which features protective back-to-back Zener diodes to prevent excessive potentials from developing in the drift gap. The drift gap was operated at a field of about 740V/cm, and the transfer and induction gaps were operated at around 3kV/cm. The GEM's each had about 450V applied across them to comfortably achieve a gain of a few thousand.

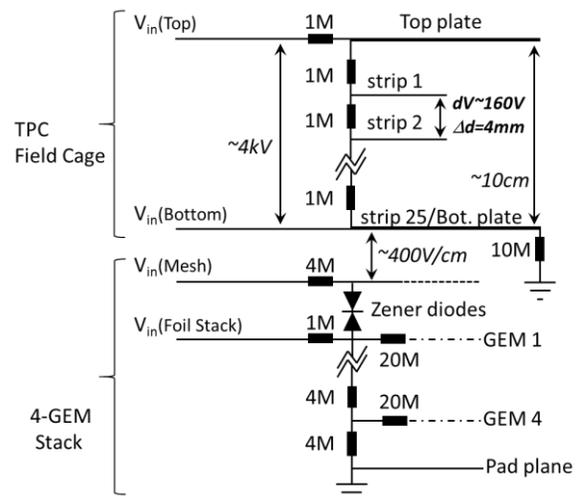

**Fig. 4 Sketch of the voltage divider for TPC field cage and the quadruple GEM beneath it.**

The degree of electric field non-uniformity within the drift volume was studied using the ANSYS finite element simulation program to calculate the vector field given the boundary conditions defined by the potentials of the field cage described above. The results from this exercise are summarized in Fig. 5 and show that the maximum deviations of the electric field components perpendicular to the drift direction are less than 0.5% of the average field over the full drift length in the fiducial volume. The regions with maximum non-uniformities tend to be close to the wire electrodes of the field cage as expected. However, for the majority of the drift volume the deviations are far less and signify that the non-uniformities of the drift field will have a relatively small impact on the quality of the track reconstruction. These results also include the effect of the potential on the planar mesh electrode from the Cherenkov detector, which is located at the $Y$-$Z$ plane at an $X$-position a few mm from the edge of the wire plane.

### B. Cherenkov detector

The Cherenkov portion of the prototype consists of a quadruple GEM stack with a CsI photocathode coating the surface of the top GEM. A finely woven planar mesh electrode with about 90% optical transparency is used to define the drift gap and a 3x3 array of square pads make up the readout plane. The photosensitive layer is placed directly in front of the transparent side of the field cage, in a windowless, proximity focused arrangement to maximize the photoelectron yield from the incident Cherenkov light. This detector configuration was also used in the Hadron Blind Detector (HBD) for the PHENIX experiment at RHIC and demonstrated excellent performance [5].

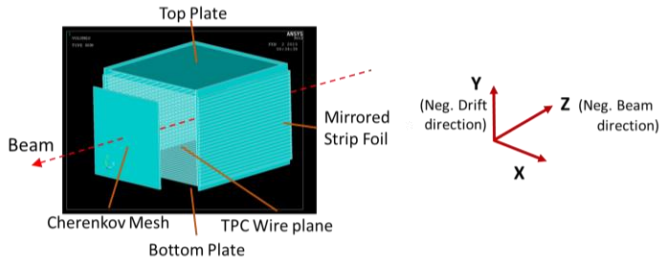

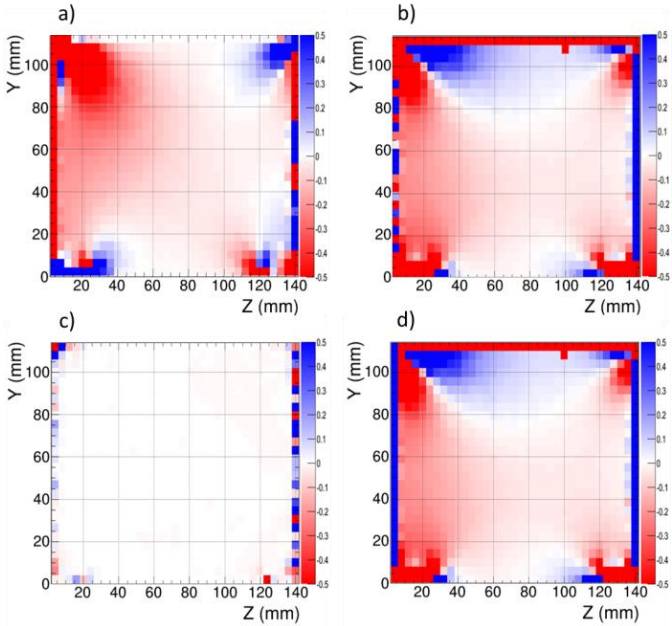

**Fig. 5** Finite element calculations of the static electric field within a cubical field cage using ANSYS. The magnitude of the vector field components are plotted in a plane that cuts the drift volume in half along the drift direction. Each component is renormalized to show the deviation from the average throughout the fiducial drift volume, along: a) *X*, b) *Y*, c) *Z*, and d) the resultant vector.

In this detector configuration, the sensitivity of CsI is nicely matched to the deep ultraviolet portion of the Cherenkov spectrum, where the intensity grows as the inverse square of the wavelength. At 200nm the CsI quantum efficiency is roughly zero, but increases to above 50% at the transmittance cutoff for $CF_4$ of approximately 112nm. Due to the resulting high figure of merit ($N_0$) for pure $CF_4$, the photon yield per cm for the Cherenkov radiator is maximized for this detector. The size of the readout pads were 3.3cm x 3.3cm so that the Cherenkov cone is mostly captured by the central pad when the beam is appropriately centered on the active area of the detector. In order to minimize absorption losses due to impurities in the gas and to keep the quantum efficiency of the CsI photocathode from degrading, the gas purity levels were maintained at the level of tens of ppm's of water and oxygen.

The quadruple GEM was powered using a similar passive voltage divider described in Fig. 4 for the TPC GEM stack. The transfer and induction fields were also similar (~3kV/cm), but the fields for the GEMs were generally higher in order to have a higher gain due to a smaller primary charge from the Cherenkov light. The drift field was operated at forward bias to bring charge deposited in the drift gap to the GEMs and also at a slight reverse bias to repel most of this charge, as discussed in later sections.

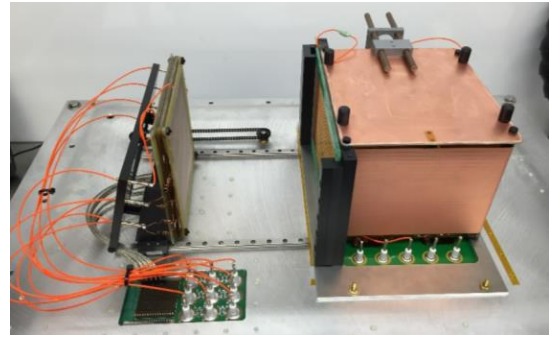

**Fig. 6** Photo of the fully assembled prototype TPCC detector.

The interior of the fully assembled detector is shown in Fig. 6. On the left, the Cherenkov detector is mounted to a rail system that allows the distance between the photocathode and the field cage to be varied. This allows the radiator length and thus the photon yield to be changed in order to conduct systematic studies discussed later. The chamber is sealed using an aluminum enclosure and a gas recirculation system maintained the gas flow while removing impurities. As shown in Fig. 7, the gas enclosure includes a beam entrance and exit window; each is made with a .002" thick mylar sheet to reduce the material budget along the particle path.

The CERN SRS system and APV25 front end cards are used to read out the charge from all the pads of both detectors [6]. This DAQ hardware features an analog charge sensitive preamplifier that generates a waveform with an 80ns rise time and is digitized by a 12-bit (although effectively 11 bits) 40MHz ADC over 27 time samples. The data is recorded using the RCDAQ data acquisition software [7] which collects the data and writes it to disc.

### III. PERFORMANCE

The TPCC prototype was studied in the FTBF by exposing it to both the primary 120GeV/c proton beam and to a secondary mixed beam consisting of electrons, pions, kaons, and protons at energies ranging from 4 to 12GeV. The prototype was also placed just downstream of a 12 layer Si telescope [8] which was used to measure reference tracks with very high resolution for comparison with tracks measured in the TPC. A photo of the TPCC prototype installed in the MT6.1A area of the FTBF is shown in Fig. 7, next to the silicon telescope to the left.

#### A. TPC detector

For the TPC, 120GeV proton tracks were reconstructed along two orthogonal planes, defined by the *X-Z* and *Y-Z* axes of the zigzag readout plane, as shown in Fig. 2. The *X*-coordinate corresponds to the position sensitive coordinate of the zigzag pattern where the hit position of impingent charge clouds are interpolated between several pads by calculating a charge weighted mean (or centroid) of the fired pads. The beam of particles entered the detector such that their trajectories were along the Z-axis. Thus, a space point was able to be computed for each pad row, along Z to establish an X, Z coordinate pair as illustrated by the left panel of Fig. 8, which shows the response of the TPC readout pads in the X-Z plane. The Y-coordinate is along the drift direction and is determined by extracting timing information from the rising edge of the waveform signal of each fired pad. The rising edge was fit to a Fermi-Dirac function whose inflection point returned the charge arrival time, $\tau$. The drift distance along Y was then computed as $v_d * \tau$, where $v_d$ is the electron drift velocity, equal to 7.5μm/ns at 400V/cm in pure $CF_4$. Accordingly, the computed drift distance for each pad and its corresponding pad row establish a Y, Z coordinate pair. In both cases, the Z coordinate is essentially a dummy variable and conveys little positional information, due to the rather coarse segmentation of the pad-rows. Ultimately, this TPC configuration provides precision 2D spatial coordinates for tracks in the X-Y plane. Once the 10 coordinate pairs for each plane are determined, the series of points are fit to a line to reconstruct the track. An example of such a linear fit is shown in the right panel of Fig. 8 for a slightly inclined track in the Y-Z plane, which was obtained by tilting the detector chamber with respect to the beam.

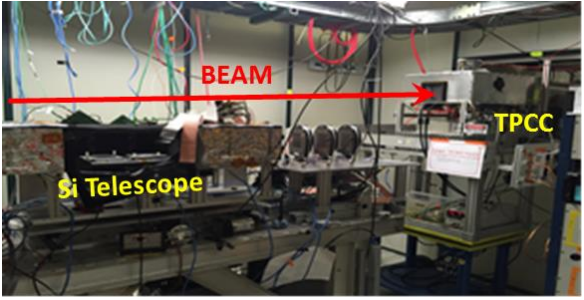

**Fig. 7** Photo of the TPCC prototype in the beamline at FTBF. The silicon tracking detector, just upstream of the prototype, is lowered out of the beam in this photo.

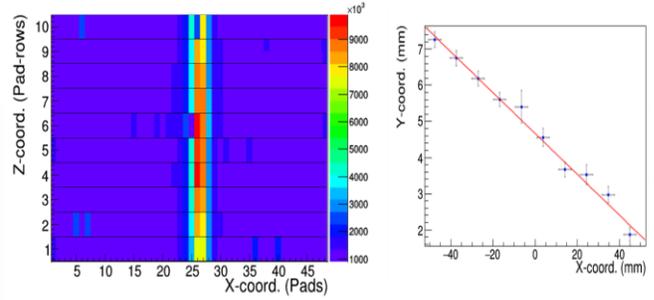

**Fig. 8** Left: TPC event display from a horizontal track, showing fired pads on the 10 pad-rows of the readout plane. Right: Linear fit to reconstructed spatial points for an inclined track at ~3 degrees.

The resulting position residual distributions and the corresponding position correlation plots are shown in Fig. 9 and 10 for both horizontal tracks parallel to the Z-axis and for tracks with a 3 degree inclination to the Z-axis, respectively. The position residual for each track is formed as the difference between position coordinates as determined by the TPC and the same coordinates as determined by the silicon telescope, but projected onto the TPC coordinate system. The width (sigma) of each residual distribution is taken to be the position resolution for the TPC detector, after the intrinsic position resolution of the silicon for projected tracks is taken into account, which is essentially negligible (i.e., 17μm for a ~1m track projection, subtracted in quadrature). The position resolution for horizontal tracks in the TPC was found to be about 80μm and 167μm along the X and Y coordinates respectively. The results at a 3 degree inclination were similar: 88μm and 151μm for X and Y respectively, where minor changes in the detector gain and the degree of transverse diffusion over slightly different drift lengths for each detector orientation could account for the small differences in resolution. The angular resolution for the track components in the X-Z and Y-Z planes were determined in a similar fashion and were found to be a little over 2mrad, respectively for each plane, as shown in Fig. 11. Since the beam at FTBF has very small divergence (150-300μrad), all the particle trajectories were considered parallel, making the width of the distributions of the reconstructed track angles effectively equivalent to the width of the corresponding residual distributions. The results for the zero degree inclination and 3 degree inclination are almost identical.

Roughly half of all the clusters of charge collected by each pad-row fired a single pad, so interpolating the hit position was not possible. This resulted in a significant degradation of the single point resolution for pad rows with only one pad firing [9]. For this reason, tracks were only reconstructed for events where at least three pad-rows consisted of 2 or more fired pads. Ultimately, this event selection scheme resulted in excluding more than a third of the events from the analysis.

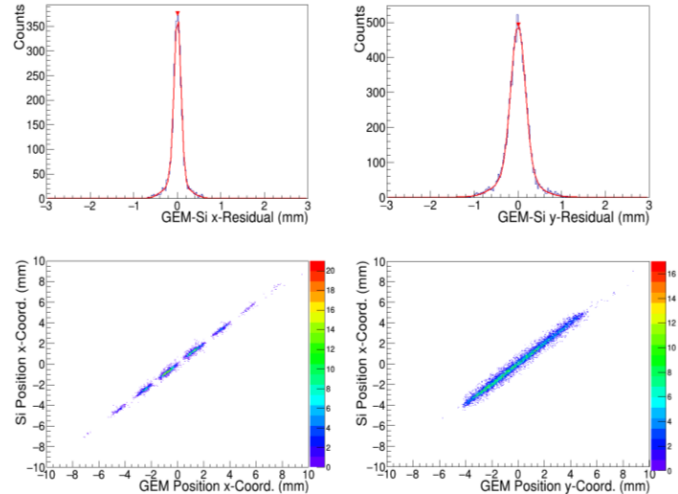

**Fig. 9** Top: residual distributions for the X and Y coordinates of reconstructed track positions in the TPC, for tracks with zero degree inclination with respect to the pad plane. Bottom: Scatter plots of the associated reconstructed track positions correlated with the results from the silicon telescope.

In contrast, the Y-position measurement is mostly unaffected by single pad clusters, since calculating the charge arrival time involves taking an average of the timing from every fired pad within a pad-row, which is not badly affected if only a single pad fires. The timing resolution for determining the Y-positon was mostly determined by the 40MHz sampling rate of the ADC, which provided just 2-4 samples on the rising edge of the waveform. Ultimately, the timing resolution was found to be slightly smaller than a single bin width (25ns).

The rather large number of single pad hits in these measurements was due to very low transverse diffusion in pure

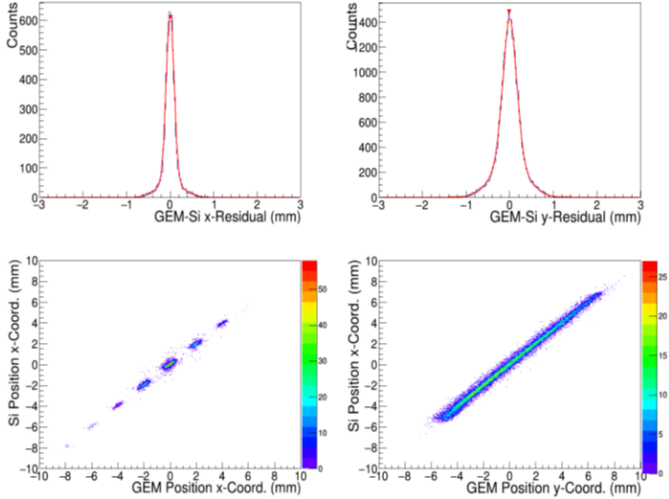

**Fig. 10** The same plots shown in Fig. 10, except for particle tracks with 3 degree inclinations with respect to the pad plane.

$CF_4$ (~122μm/$\sqrt{cm}$ at 400V/cm), which is responsible for small charge cloud sizes that are not particularly suited for the pitch of this readout. This is exacerbated by the relatively large region near the center of each zigzag pad (corresponding to 60% of the pitch) where there is no overlap with adjoining pads to enable charge sharing. As mentioned previously, the resulting single pad hits are removed from the analysis, which represents an efficiency loss in both the number of detected events and in terms of effective dead areas on the readout where there is virtually no positional sensitivity. These dead areas are apparent in the *X*-position correlation plots of Figs. 9 and 10 where the gaps in each scatter plot are associated with regions of the readout that coincide with the center of each pad.

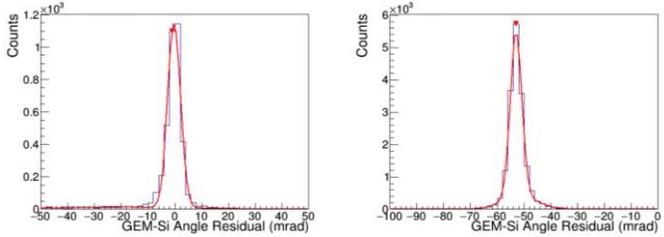

**Fig. 11** Distributions of reconstructed track angles for tracks with a 3 degree inclination with respect to the pad plane. In the *X-Z* plane the angle of incidence is zero degrees.

The transverse diffusion in this detector may be improved by using alternate gases or by reconfiguring the fields in the transfer gaps, although these approaches will likely involve compromising important gas characteristics like the photon yield or the charge transfer efficiency. However, if the zigzag pattern design is also improved to maximize charge sharing, an appropriate gas mixture may be chosen to adequately satisfy all detector requirements. In the time since these measurements were taken, we have in fact significantly optimized the design and performance of the zigzag readout board such that no single pad clusters are observed [4]. In addition, biases in charge sharing which lead to deviations from a linear response (known as a differential non-linearity) have been strongly suppressed with newer, optimized zigzag designs. Both advances have substantially improved the performance of the readout board which potentially can greatly benefit the TPC portion of the TPCC detector.

The residual distributions for each coordinate were fit to a double Gaussian function, with a dominant background component. The background is seen in the long tails of each residual distribution, which are non-negligible. However, it has been found that these background components are mostly correlated with small blocks of events in the TPCC data which have become de-synchronized with respect to the silicon tracker data. As a result, events in the background generally correspond to random residuals and do not reflect a legitimate detector response, which in general are difficult to eliminate on an event by event basis. For this reason, the detector resolution is quoted as the width of the dominant Gaussian component only.

Lastly, though the full drift time for the primary charge generated by incident tracks is several microseconds before it is collected by the readout, the capture window for the DAQ starts accepting data after a delay is applied with respect to the global trigger. This way all the charge from horizontal tracks, as well as slightly inclined ones may be acquired over the limited 700ns time window of the DAQ hardware (i.e., 28 samples*25ns time bins).

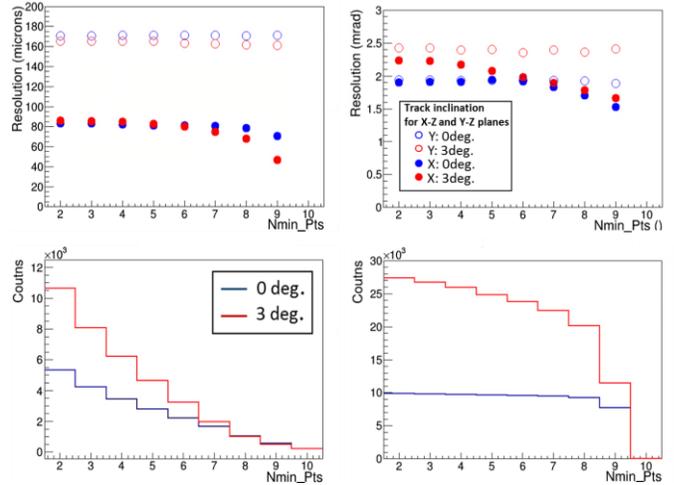

**Fig. 12** Top: position and angular resolution vs the minimum number of space points used to reconstruct each track vector; left: *X*-coordinate, right: *Y*-coordinate. Bottom: the associated number of events with the minimum number of space points used for reconstruction.

Since the detector performance depends on the number of space points used to define each particle track, a study was also undertaken to understand how the position and angular resolutions change as a function of the number of space points. The plots in Fig. 12 show the resulting position and angular resolutions if a minimum number of space points are used. As described above, the number of space points varies according to the number of pad rows removed from the analysis due to the presence of single pad clusters. The number of events as a function of the minimum number of space points is also shown in the figure, where there is a considerable drop as more pad-rows are used for the *X*-coordinate. Hence, the improvement in the resolution with more available space points comes at the cost of significantly lowering the efficiency for high quality track vectors. Conversely, the number of events stays relatively flat for the *Y*-coordinate, which is due to the fact that single pad hits do not affect the estimate of the *Y*-position (time coordinate) for each pad-row very much. As a result, there is virtually no change in the *Y*-position resolution, as expected.

*B. Cherenkov detector*

The Cherenkov detector is operated in a threshold mode such that its sensitivity to hadrons is minimized while the efficiency for detecting particles above the Cherenkov threshold is maximized. As high energy particles emitting Cherenkov light enter the TPCC chamber, they traverse an effective radiator length defined by a path through the field cage and the distance between the wire plane and the mesh of the Cherenkov GEM detector, shown in Fig. 1. The expression for the expected photoelectron signal is given by:

$$N_{pe} = N_0 \, L <sin^2\theta> = L \int_{100nm}^{200nm} Y_{Ch}(\lambda) \, QE_{CsI}(\lambda) \, T(\lambda) \, \varepsilon_C(\lambda) \, d\lambda,$$

where $N_0$ is the quality factor of the detector, which incorporates the spectral response of the detector as well as various efficiencies; $L$ is the effective path length of the radiator, $\theta$ is the Cherenkov angle, $Y_{Ch}(\lambda)$ is the Cherenkov yield per unit length, $QE_{CsI}(\lambda)$ is the photocathode quantum efficiency, $T(\lambda)$ is the combined transparency of the gas, GEM mesh electrode and GEM foil, and $\varepsilon_C(\lambda)$ incorporates the various photoelectron collection efficiency losses at the level of the GEM readout, including the transport and extraction [10]. However, since the measured signal is directly proportional to the primary number of electrons detected, a practical expression for the primary charge in this application is as follows:

$$N_e \approx N_e^{mip1}\varepsilon_{CE}^{mip1} + N_e^{mip2}\varepsilon_{CE}^{mip2} + N_e^{mip3}\varepsilon_{CE}^{mip3}G_{GEM}^{-1/4} + N_{pe}\varepsilon_{CE}^{pe}.$$

To get an accurate account of the primary number of photoelectrons, which are relatively few in number, every effort must be made to separate out the different components of the primary signal. The observed signal is thus broken down into the primary charge from ionization in the various gaps of the GEM stack, given by $N_e^{mip}$ (illustrated in Fig. 13) and the primary number of Cherenkov photoelectrons generated at the photocathode surface, given by $N_{pe}$, where $\varepsilon$ represents the associated charge collection efficiencies, and $G_{GEM}$ is the gain of the top GEM foil.

The charge from ionization may be estimated by dividing the total energy deposited in each gap of the GEM stack (shown in Fig. 13) by the charge required to create a single electron-ion pair. A minimum ionizing particle in $CF_4$ deposits an amount of energy in the gas according to: $dE/dx$~7keV/cm [11], and the ionization potential for creating an electron ion pair in $CF_4$ is 54eV [11]. As a result, about 30 electrons are deposited in the 2.3mm drift gap due to ionization. Similarly, about 20 electrons are liberated in the first transfer gap, which is about 1.6mm wide. However, since the transfer gap electrons do not undergo multiplication by the first foil, this signal is diminished by the corresponding gain (estimated to be the fourth root of the total GEM gain), making the effective contribution from this gap equal to only a few electrons for a gain of a few thousand. (The effective charge in the remaining gaps are considered negligibly small and are ignored.) The remaining part of the detected signal is from the primary photoelectrons from the Cherenkov light produced plus the scintillation photons produced by the charged track in CF4. However, since the distribution of scintillation photons is isotropic, their contribution in this configuration is very small [12]. Therefore, the corresponding photoelectrons are simply considered to be a part of the overall primary photoelectron signal.

To achieve the highest efficiency for electron identification (eID), the photoelectron collection efficiency must be maximized while the collection of the charge from hadrons must be suppressed. This is accomplished by tuning the bias field in the drift gap appropriately [11]. By operating the drift field in a forward bias mode (~1.0kV/cm), all the charge deposited in the drift gap, including that from ionization as well as the Cherenkov light is collected. However, at an optimized reverse bias field, most of the hadronic signal is repelled, while the photoelectron collection efficiency remains high [10].

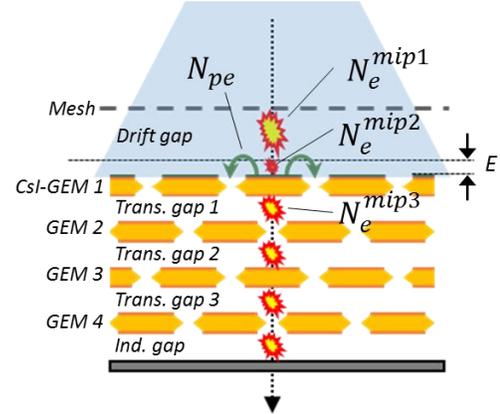

**Fig. 13 Charge deposited in Cherenkov GEM detector by an incident charged particle. The charge due to ionization is depicted by the bursts outlined in red and the green arrows show the primary Cherenkov photoelectrons extracted from the CsI photocathode and driven into the holes of the top GEM foil by the curling fringe field near the GEM surface. "E" denotes an exclusion zone for the charge due to ionization, described in the text.**

In order to study this effect, the drift field was scanned during the beam test and the results are presented in Fig. 14, which shows the mean of the summed signals from all the fired pads of the Cherenkov readout. At reverse bias fields, the majority of the charge from ionization is carried away from the GEM detector and towards the mesh. However, a portion of this charge (about 2 electrons) in the so-called exclusion zone region of the drift gap (~100μm above the top GEM surface) [11] experience roughly the same collection efficiency as the released photoelectrons near the GEM surface. The effective photoelectron yield at reverse bias is then roughly the sum of the last three terms in the expression for $N_e$. Likewise, the residual signal from charged particles with energies below the Cherenkov threshold originate from the middle two terms.

The ratio of the effective photoelectron signal to the residual charge from ionization corresponds to how well the electron and hadron signals are separated for efficient eID [11], which clearly improves in proportion to the true photoelectron yield.

It has been found that the optimum field for maximizing the eID for this detector configuration is at around -50V/cm [10]. The difference in signal size at 300V/cm and at 0V/cm corresponds to $N_e^{mip1}$, which may be estimated in terms of absolute charge to provide a means to calibrate the ADC scale. Thus, with prior knowledge of the absolute detector gain and the value of the collection efficiencies involved, the expression for $Ne$ may be used to estimate the effective photoelectron yield at the optimum reverse bias field, where photoelectron collection is maximum and $N_e^{mip1} = 0$. This number was found to be $11 \pm 2$ photoelectrons, in good agreement with the expected number, extrapolated from the HBD results in [5]. The total remaining primary signal from ionization was determined to be around $4 \pm 1$ electrons.

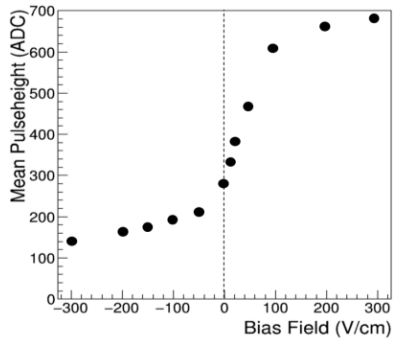

**Fig. 14 Mean signal of signal from Cherenkov detector Vs bias field applied to the drift gap.**

Another method was employed to determine the effective photoelectron yield as a function of the radiator length. In this case, the signal from the Cherenkov detector was measured at successive positions of the photosensitive GEM on its rail, which defines the effective length of the radiator. The results of this measurement are shown below in Fig. 15 at the optimized reverse bias field. Like the plot in Fig. 14, this plot also has a built in means for calibrating the ADC scale. In this case, the horizontal dashed line corresponds to the total ionization charge collected at reverse bias, found to be about 4 electrons, making the photoelectron yield at 29 cm about $12 \pm 3$ photoelectrons, in rough agreement with the earlier results. Though more sophisticated methods exist [12] for precisely determining photoelectron yields, we have adopted the straight forward approach above which is adequate for the purposes of demonstrating a proof of principle.

For the purpose of determining the eID performance, the detector was exposed to a 12 GeV beam of mixed particles, consisting of electrons, pions, kaons, and protons. A differential Cherenkov counter, also in the beam just upstream from the TPCC, was tuned to provide a trigger for the lighter electrons and pions. This provided a way to tag events in the TPCC corresponding to electrons and pions (e/π) which generate Cherenkov light at this beam energy. The pulse height distribution resulting from these tagged events with the detector operated at three different gains is shown in Fig16, where the gain was divided out from the total charge to reveal the primary charge. The Cherenkov trigger was also used as a veto to generate similar pulse height distributions from the hadrons (K/p) that do not generate light, also shown in Fig.16.

The sensitivity to hadrons is mostly due to the charge deposited in the first transfer gap. Therefore, the relative magnitude of this signal with respect to the Cherenkov signal is diminished as the total GEM gain is increased. This can be seen as the pulse height distributions from the hadrons in the beam become more separated from the e/π distribution at higher gain. Accordingly, the Cherenkov detector may be a highly effective eID detector if the gain is turned up sufficiently high and the

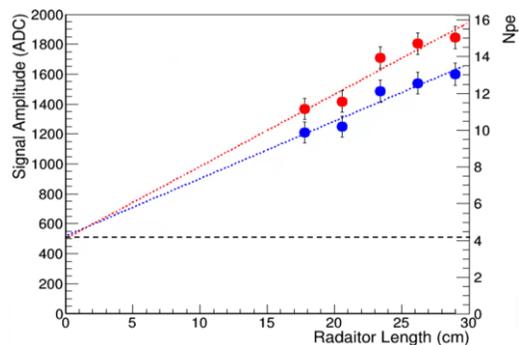

**Fig. 15 Mean Cherenkov signal amplitude (from the sum of all fired pads and the central pad only) vs the effective length of the radiator. The horizontal dashed line corresponds to the extrapolated signal where the radiator length is zero and no Cherenkov light is produced.**

$N_{pe}$ threshold is set appropriately. The measurements here could not be performed at higher gains since the front end electronics would saturate. However, it is obvious that the eID performance may be considerably improved if the top GEM foil gain were increased relative to the bottom three foils or if a significantly longer radiator were employed.

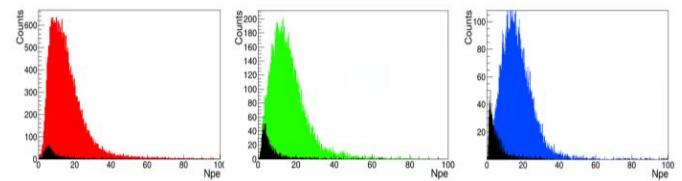

**Fig. 16 Measured pulse height distributions generated from a e/π trigger at three different detector gains: red ~1000, green: ~3300, blue: 6000. The black histogram represents the corresponding hadron (K/p) signal at the same gain.**

The plot in Fig. 17 demonstrates the improvement in the eID efficiency both as the gain is increased and as a function of the photoelectron threshold. At low gain, the relative eID efficiency, taken as the ratio of all the Cherenkov triggers recorded during a run to the total number of triggers, stays flat as a function of $N_{pe}$ and reveals the e/π fraction present in the beam (~90%). At higher gains, and as the $N_{pe}$ threshold is increased, this relative measure of the efficiency goes above the nominal make up of e/π in the beam, by up to 8%. Ultimately, the eID performance is limited by the sensitivity to the long Landau tails of the MIP signal in the first transfer gap. [5].

### C. TPC – Cherenkov detector correlations

So far the two component detectors of the TPCC have been shown to work successfully in standalone mode. However, to fully validate this detector concept, it must be shown that the performance of one detector is not compromised by the presence of the other. More specifically, since the Cherenkov mesh operates at about 4kV and is parallel to the wire plane of the TPC field cage, which operates between 4-8kV, there exists the possibility that the drift field will be slightly distorted when the two planes come into close proximity. In this case the angular resolution was used as a probe to signify any changes in the field uniformity as the Cherenkov mesh was brought closer to the field cage, under the assumption that the resolution would quickly degrade if there is any influence from the mesh electrode on the drift field. The results are shown in Fig. 18, and basically reveal no significant effect at even a ~1cm separation between the Cherenkov mesh and the wire plane.

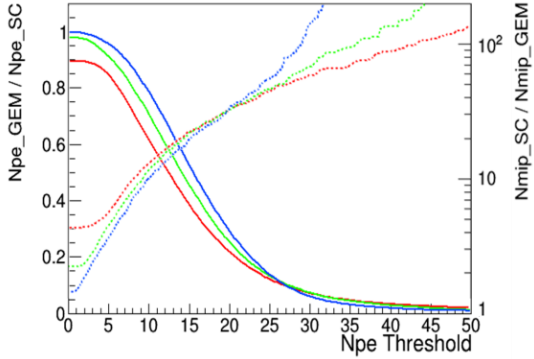

**Fig. 17** The number of detected Cherenkov triggers (Npe_GEM) and hadron triggers (Nmip_GEM), divided by the total number of Cherenkov triggers measured by the Cherenkov counter (Npe_SC) and the number of hadrons measured by a scintillation counter respectively, as a function of $N_{pe}$, at the three different values of gain: red ~1000, green: ~3300, blue: 6000.

Finally, the correlation observed for the *X* and *Y* coordinates of each track measured by the two detector components shows both detectors were responding to the same particles entering the detector chamber. The hit position of each particle track was estimated in the Cherenkov detector by using a weighted mean of the Cherenkov signal on the readout plane.

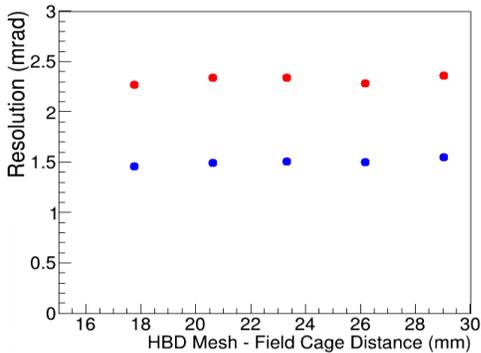

**Fig. 18** Angular resolution of TPC tracks vs distance between the wire frame of the field cage and the Cherenkov mesh.

Although the resulting hit position had a relatively poor resolution due to the coarse segmentation of the pads, a relatively strong hit correlation was also observed with the track positions reconstructed in the TPC, as shown in Fig. 19.

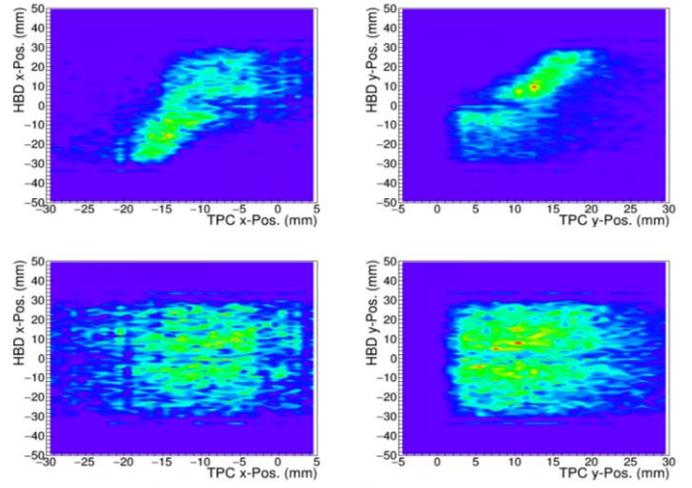

**Fig. 19** Top: Correlation of *X* and *Y* hit positions of particle tracks measured in the TPC and the Cherenkov detectors, respectively. Bottom: For comparison, the events in the two detectors are purposefully randomized to demonstrate the case with no correlation.

## IV. SUMMARY

A combined TPC/Cherenkov prototype detector has been developed to provide tracking information and particle identification within a common detector volume. The detector uses ionization for tracking and UV light from particles above the Cherenkov threshold for pID. In addition, *dE/dx* information from the TPC could also be used for additional pID. The prototype was tested at the Fermilab test beam facility to provide a proof of principle for this hybrid detector concept and showed very good tracking and eID performance. In addition, the two different detector technologies employed were shown to work together in a complimentary way without imposing limiting factors on one another. Such a detector could enhance the capabilities of a central TPC tracker for an Electron Ion Collider detector by helping to identify the scattered electron in ep and eA collisions in the central region by combining the tracking and pID capabilities into a single detector.

## V. ACKNOWLEDGEMENTS

We would like to acknowledge the following Stony Brook undergraduate students for their help in preparing the detector apparatus and their contributions toward data taking shifts: B. Kestelmann, L. de Bruin, H. Sun, O. Park, and E. Samoylovich.

## VI. REFERENCES

[1] E.C. Aschenauer et al., "eRHIC Design Study: An Electron-Ion Collider at BNL", arXiv:1409.1633 (Dec 2014)

[2] C. Woody et al, "A Prototype Combination TPC Cherenkov Detector with GEM Readout for Tracking and Particle Identification and its Potential Use at an Electron Ion Collider", Proceedings of the 2015 Micropattern Gas Detector Conference, Trieste, Italy, October 12-15, 2015.

[3] B. Yu, et al., "Study of GEM Characterisitcs for Application in a Micro TPC," IEEE Transactions on Nuclear Science, Vol. 50, No. 4, August 2003.

[4] B. Azmoun, et al., "Design Studies for a TPC Readout Plane using Zigzag Patterns with Multistage GEM Detectors", IEEE Trans. Nucl. Sci. Vol 65, No.7 (2018) 1416-1423.

[5] W. Anderson et al., "Design, construction, operation and performance of a Hadron Blind Detector for the PHENIX experiment", Nucl. Inst. Meth. A646 (2011), pp. 35-58.

[6] S. Martoiu, et al, "Development of the scalable readout system for micro-pattern gas detectors and other applications" JINST 8 C03015, March 2013.

[7] M. L. Purschke, "Readout of GEM stacks with the CERN SRS system", Real Time Conference (RT) 2012 18th IEEE-NPSS, pp. -3, 2012.

[8] S. Kwan, et al, "The CMS Pixel Tracking Telescope at the Fermilab Test Beam Facility", FERMILAB-PUB-14-212-CDM.

[9] B. Azmoun, et al., "A Study of a Mini-drift GEM tracking detector," IEEE Transactions on Nuclear Science, Vol. 63, No. 3 June 2016.

[10] B. Azmoun et al, "Collection of Photoelectrons and Operating Parameters of CsI Photocathode GEM Detectors", IEEE Transactions on Nuclear Science, Vol. 56, No. 3, June 2009.

[11] Z. Fraenkel et al., "A Hadron Blind Detector for the PHENIX experiment at RHIC", Nucl. Inst. Meth. A546 (2005), pp. 466-480.

[12] B. Azmoun, et al, "A Measurement of the Scintillation Light Yield in Using a Photosensitive GEM Detector", IEEE Transactions on Nuclear Science, Vol. 57, No. 4, August 2010.